\documentstyle[preprint,aps,psfig]{revtex}

\def\lsim{\raise0.3ex\hbox{$\;<$\kern-0.75em\raise-1.1ex
\hbox{$\sim\;$}}}
\def\gsim{\raise0.3ex\hbox{$\;>$\kern-0.75em\raise-1.1ex
\hbox{$\sim\;$}}}
\begin{document}

%%%%%%%%%%%%%%%%%%%  title page %%%%%%%%%%%%%%%%%%%%%%%%%%%%
\baselineskip 8.2mm
\begin{flushright}
TMUP-HEL-0017 \\ 
hep-ph/0009019 \\ 
 \end{flushright}
\vspace{-1.0cm}
\begin{center}
\Large\bf
A Two-Dimensional Model with Chiral Condensates and 
Cooper Pairs Having QCD-like Phase Structure
\end{center}
\begin{center}
Alan Chodos\footnote[1]{E-mail: chodos@aps.org}\\
{\it  Department of Physics, Yale University, New
Haven, CT 06520-8120, USA  and  American Physical Society, One
Physics Ellipse, College Park, MD 20740-3844 USA}\\
\vskip 0.3cm
Hisakazu Minakata\footnote[2]{E-mail: minakata@phys.metro-u.ac.jp}\\
{\it Department of Physics, Tokyo Metropolitan University \\
1-1 Minami-Osawa, Hachioji, Tokyo 192-0397, Japan, and \\
Research Center for Cosmic Neutrinos, 
Institute for Cosmic Ray Research, \\ 
University of Tokyo, Kashiwa, Chiba 277-8582, Japan}
\vskip 0.3cm
Fred Cooper and Anupam Singh\footnote[3]
{E-mail: cooper@schwinger.lanl.gov, singh@lanl.gov}\\
{\it Los Alamos National Laboratory, Los Alamos, NM 87545, USA}
\vskip 0.3cm
Wenjin Mao\footnote[4]{E-mail: maow@physics.bc.edu}\\ 
{\it Department of Physics, Boston College, Chestnut Hill, MA 02467, USA}
(September 2000) 
\end{center}
\vspace{-1.0cm}
\begin{abstract}
We describe how a generalization of the original Gross-Neveu model
from $U(N)$ to $O(N)$ flavor symmetry leads to the appearance of a
pairing condensate at high density, in agreement with the
conjectured phenomenon of color superconductivity in
$(3+1)$-dimensional QCD.
Moreover, the model displays a rich phase structure which closely 
resembles the one expected in two-flavor QCD.
\end{abstract}
\newpage

\section{Introduction}

The work that will be described in this note had its origin at
the beginning of the history of TMU-Yale collaboration about a
decade ago, when one of us (A.C.) was working on QED with a  chemical
potential \cite{ref1,ref2} and another (H.M.) suggested that the study 
of the Gross-Neveu model \cite{ref3} with a chemical potential might 
have interesting applications to doped polyacetylene. This resulted 
in a pair of papers, in the first of which \cite{ref4} the effective 
potential for the Gross-Neveu model was computed with a non-vanishing 
chemical potential $\mu$ at leading order in large $N$. In agreement 
with earlier work \cite{ref5}, a first-order phase transition was found 
at $\mu = m/\sqrt{2}$, where $m$ is the fermion mass, and this result 
was successfully compared to the phenomenology of doped polyacetylene 
\cite {ref4,mina1}. 
In the second paper \cite{ref6}, the thermodynamic Bethe ansatz was 
employed to compute the next-to-leading $1/N$ correction to the 
position of the phase-transition.

But the Gross-Neveu model, possessing aymptotic freedom and
spontaneous chiral symmetry breaking, was originally introduced as
an analogue of QCD. With renewed interest in QCD at finite density
\cite{ref7}, the TMU-Yale collaboration shifted its focus to trying
to understand the phenomenon of color superconductivity. Before
describing that work, we make a few remarks about the role played
by the large $N$ approximation, which will be used throughout this
paper, in the prediction of the superconducting phase transition.

The issue is the following: ~the superconducting phase will involve
a non-vanishing pairing condensate, which spontaneously breaks the
$U(1)$ of fermion number. But the generalized Gross-Neveu model is 
a $1 + 1$ dimensional theory, and theorems exist \cite{ref8} that 
forbid the spontaneous breaking of continuous symmetries in $1 + 1$
dimensions. Nevertheless, to leading order in $1/N$ one can compute
the effective potential, and one finds that the condensate indeed
forms. As Witten has explained \cite{ref9}, corrections are
expected beyond the leading order which will remove the condensate.
More explicitly, the vacuum expectation value of $B(x)$ must 
vanish at any finite $N$ because its correlation function 
$\langle B(x) B^{\dagger}(y)\rangle$ will fall off, for
large $(x - y)$, as $\mid x - y \mid^{- c/N}$. 
If we take the limit $N \rightarrow \infty$ first before taking 
the limit $\mid x - y \mid \rightarrow \infty$ it implies 
non-vanishing vacuum expectation value of $B(x)$. 

One can enhance one's intuition on this matter if one accepts the
analogy between the infinite number of fields that are present in
the large $N$ approximation and the infinite set of fields one
obtains upon making a Kaluza-Klein reduction from a higher number
of dimensions. Admittedly, this analogy is not perfect, because in
the large $N$ approximation the fields are all identical, whereas
in Kaluza-Klein they typically represent a tower of states with
indefinitely increasing masses. Nevertheless, the analogy suggests
that the leading $N$ theory somehow shares the physics of a theory
in a higher number of dimensions, where the no-go theorems do not
apply. The analogy further suggests that if one is interested
primarily in elucidating properties of $(3 + 1)$-dimensional QCD,
the large $N$ approximation to the Gross-Neveu model may be a more
appropriate object of study than the full theory.

Even more relevant examples in this context are provided by the 
Eguchi-Kawai model \cite {EK82}, or by a class of large-N matrix 
models \cite {IKKT,BFSS}. In these matrix models the eigenvalues 
of the matrix are somehow converted, in the large-N limit, 
into the spatial dimensions. Thus, the matrix models in zero dimension 
thereby serve as definitions of 4-dimensional Yang-Mills and 
higher-dimensional superstring or supergravity theories. 
It is not clear to us if the analogy goes through to the vector-like 
large $N$ theories such as ours. 
Nonetheless, it is at least suggestive enough to lead to the 
speculation that Coleman's theorem is violated in $(1 + 1)$-dimensional 
large $N$ theories because they are actually the theories in higher 
dimensions.

\section{The $O(N)$ Gross-Neveu Model}

The phase structure of the original Gross-Neveu model is well-known
\cite{ref5}: ~at low temperature and density, the chiral symmetry
is broken; as either is increased, the chiral condensate vanishes
and the symmetry is restored. But there is no evidence for a
second, pairing condensate that would be the analogue of color
superconductivity in QCD. It is necessary to add to the original
model an extra term in the Lagrangian which, as it turns out, has a
simple interpretation. The original model, with $N$ identical fermi
fields, possesses $U(N)$ symmetry. The extra term is what can be
added if one demands only $O(N)$ symmetry in flavor space.

Explicitly, our lagrange density reads

\begin{eqnarray}
{\cal L} &=& \bar{\psi}^{(i)} i \bigtriangledown  \!\!\!\!\!\! /
\psi^{(i)} + {1
\over 2} g^2 [\bar{\psi}^{(i)} \psi^{(i)}][\bar{\psi}^{(j)} \psi^{(j)}]
\nonumber \\
&+& 2 G^2 (\bar{\psi}^{(i)} \gamma_5 \psi^{(j)})(\bar{\psi}^{(i)}
\gamma_5 \psi^{(j)}) - \mu \psi^{\dagger (i)} \psi^{(i)} .
\end{eqnarray}

\bigskip\noindent
The flavor indices $i, j$ are summed from $1$ to $N$. The first $2$
terms are the original $GN$ model. The third term, because of the
way the flavor indices are summed, is $O(N)$ but not $U(N)$
invariant. The last term represents the chemical potential. The
gamma matrices are $2 \times 2$, and we choose $\gamma^0 =
\sigma_1, \gamma^1 = - i\sigma_2; \gamma_5 = \sigma_3$. Then

\begin{eqnarray}
2G^2 \bar{\psi}^{(i)} \gamma_5 \psi^{(j)} \bar{\psi}^{(i)} \gamma_5
\psi^{(j)} = - G^2 [\epsilon_{\alpha\beta} \psi_{\alpha}^{\dagger (i)}
\psi_{\beta}^{\dagger (i)}]
[\epsilon_{\gamma \delta} \psi_{\gamma}^{(j)}
\psi_{\delta}^{(j)}] ~.
\end{eqnarray}

\bigskip\noindent
Following standard techniques \cite{ref10} we introduce auxiliary
fields:

\begin{eqnarray}
m(x) &=& - g^2 \bar{\psi} \psi \nonumber \\ B(x) &=& - G^2
\epsilon_{\alpha\beta} \psi_{\alpha}^{(i)}\psi_{\beta}^{(i)}
\end{eqnarray}

\bigskip\noindent
and integrate out the fermions to obtain the effective action
$\Gamma_{eff}$ as a function of $m$ and $B$. This is described in
more detail in refs. \cite{ref11,ref12}.

As appropriate to a vacuum solution, we assume that $m$ and $B$ are
constants, and write

\begin{eqnarray}
\Gamma_{eff} = - N \int (d^2x) V_{eff} ~.
\end{eqnarray}

\bigskip\noindent
We set $g^2N = \lambda$, and $G^2 N = \kappa/4$. The large $N$
limit is defined by letting $N \rightarrow \infty$ with $\lambda$
and $\kappa$ fixed. We find

\begin{eqnarray}
V_{eff} (m, M) = {m^2 \over 2 \lambda} + {M^2 \over 2 \kappa} +
V_{eff}^{(1)}(m,M)
\end{eqnarray}

\bigskip\noindent
where $M^2 = 4B^{\dagger}B$. Explicit computation yields
\cite{ref12}

\begin{eqnarray}
V_{eff}^{(1)} (m,M) = -{1 \over 2 \pi} \int_0^{\Lambda} dk [k_+ +
k_- + {2
\over \beta} ln (1 + e^{-\beta k_+}) + {2 \over \beta} ln (1 +
e^{-\beta k_-}) ]
\end{eqnarray}

\bigskip\noindent
where $\Lambda$ is an ultraviolet cutoff, $\beta = 1/kT$ (T =
temperature) and

\begin{eqnarray}
k_{\pm} = \sqrt{b_1 \pm 2 b_2}
\end{eqnarray}

\noindent
with

\begin{eqnarray}
b_1 = M^2 + m^2 + \mu^2 + k^2
\end{eqnarray}

\noindent
and

\begin{eqnarray}
b_2 = [M^2 m^2 + \mu^2 (k^2 + m^2)]^{1/2} ~.
\end{eqnarray}

\section{Renormalization}

As with the original GN model, we renormalize the theory, thereby
absorbing the dependence on $\Lambda$ in the relation between the
bare couplings $\lambda$ and $\kappa$ and their renormalized
counterparts $\lambda_R$ and $\kappa_R$. Recall that in the
original model, one trades the dependence on $\Lambda$ for
dependence on a renormalization scale $m_0$, or equivalently on the
fermion mass $m_F$. Once one fixes $m_F$ at its physical value,
there are no further parameters in the theory: ~the renormalized
coupling $\lambda_R$ varies with $m_0$ in such a way as to keep
$m_F$ fixed.

Similar features appear in the more general model, except that now
there are two renormalized couplings. As a consequence, the theory
has two physical parameters: ~the fermion mass $m_F$, and another
parameter $\delta$ that is independent of the choice of
renormalization scale, and is defined by

\begin{eqnarray}
\delta = {1 \over \kappa_R} - {1 \over 2\lambda_R} = {1 \over \kappa} - {1 \over 2\lambda} ~.
\end{eqnarray}

\bigskip\noindent
(Although $\kappa$ and $\lambda$ separately depend on the cutoff
$\Lambda$, the combination represented by $\delta$ does not.)
Further computation \cite{ref12} allows us to express the
renormalized $V_{eff}$ as a function of the condensates $m$ and
$M$, the physical parameters $m_F$ and $\delta$, and the
temperature $T$ and chemical potential $\mu$:

\begin{eqnarray}
V_{eff} = \delta  M^2 &-& {1 \over 2 \pi} \int_0^{\infty} dk [k_+ +
k_- + {2 \over \beta} (\ln ~(1 + e^{- \beta k_+}) + \ln ~(1 + e^{-
\beta k_-})) - 2 k_1 \nonumber \\ &-& 2 k (m^2 + M^2) {1 \over
\sqrt{k^2 + m_F^2}}] ~.
\end{eqnarray}

\bigskip\noindent
This expression is valid for $\delta > 0$. For $\delta < 0$, there
is no fermion mass, and instead at $\mu = T = 0$ there is a pairing
condensate $M^2 = \Delta^2$. Since QCD has broken chiral symmetry
but no pairing condensate at $\mu = T = 0$, we shall restrict
ourselves to $\delta > 0$ henceforth.   By studying how $\lambda_R$
and $\kappa_R$ vary with renormalization scale, one sees that both
are asymptotically free. (Because $\delta$ is independent of this
scale, it follows that if one of these couplings is asymptotically
free so must the other be.)

\section{Phase structure}

The phase structure of the theory can now be mapped out by fixing
$m_F, \delta, \mu$ and $T$, and finding the minimum of $V_{eff}$
with respect to $m$ and $M$. If $m = M = 0$, the theory is fully
symmetric; if $m \neq 0$, chiral symmetry is spontaneously broken,
while if $M \neq 0$, there is a pairing condensate which represents
the analogue of color superconductivity.

Because of the complexity of $V_{eff}$, this analysis must be done
numerically. The results are schematically displayed in Figure 1 
for a typical value of $\delta$.

\vskip 2cm
\setlength{\unitlength}{0.6mm}
\begin{figure}[h]
\begin{center}
\begin{picture}(140,140)
\put(10,10){\vector(1,0){130}}
\put(10,10){\vector(0,1){130}}
\put(3,3){\large $0$}\put(57,2){\large $P$}
\put(135,3){\large $\mu$}\put(2,135){\large $T$}
\qbezier[300](60,10)(60,30)(50,60)
\qbezier[30](10,100)(40,85)(50,60)
\put(47,63){\line(1,-1){6}}
\put(47,57){\line(1,1){6}}
%\multiput(54,48)(3,0){28}{\line(1,0){2}}
\qbezier[50](54,48)(94,48)(134,48)
\put(20,30){\large $M= 0$}
\put(20,40){\large $m\ne 0$}
\put(75,30){\large $m=0$}
\put(100,30){\large $M\ne 0$}
\put(60,100){\large $m=M=0$}
\end{picture}
\end{center}
\caption{A schematic illustration of the phase diagram on $T-\mu$ 
plane.}\label{fig1}
\end{figure}
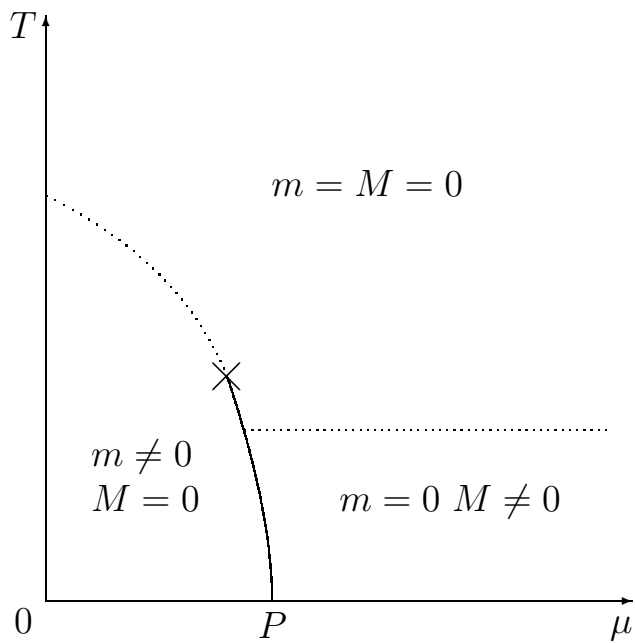

The point $\times$ is a tricritical point, at which the order of the
phase transition changes from first (solid line) to second (dotted
line). The point $P$ is located at $\mu^2 = {m_F^2 \over 2} [1 -
e^{-4\pi\delta}]$. This generalizes the earlier result for the
$U(N)$ GN model, $\mu^2 = {1 \over 2} m_F^2$, for which $\delta
\rightarrow \infty$. Note that, interestingly, even though we have 
asymptotic freedom, the
$M \neq 0$ phase persists for arbitrarily large $\mu$. In fact, the
transition temperature from the Cooper pairing phase to the
unbroken phase is independent of $\mu$.

The figure is qualitatively valid for

\begin{eqnarray}
\delta > \delta_c \cong {1.13 \over 4\pi} ~.
\end{eqnarray}

\bigskip\noindent
For smaller, positive $\delta$, the tricritical point disappears
and the phase transition for fixed $\mu$ is always second order.
For $\delta < 0$ the chirally broken phase disappears, and there is
only a pairing phase at low $T$ and an unbroken phase at high $T$.

There is no case of a phase in which both $m$ and $M$ are non-zero.
However, there can be coexistence of the $m \neq 0$ and $M\neq 0$ 
phases along the line of first-order phase transitions.

\section{Conclusions}

The phase structure displayed in Figure 1 is strikingly similar to
that conjectured in the literature for two-flavor QCD. This gives
us confidence that further properties of QCD can perhaps be studied
reliably in this $1 + 1$-dimensional theory, whereas calculations
involving the full $3 + 1$-dimensional QCD may be prohibitively
difficult. For example, work is in progress \cite{ref13} on
dynamical calculations, in which the system is prepared initially
in a dense, hot state and allowed to cool and expand. Various
distributions, such as the "pion" wave function and the "pion"
correlation function, are followed as this evolution takes place.
So far these computations have been performed only in the $U(N)$
model, and in the approximation that the expectation value of
physical quantities depends only on proper time $\tau = \sqrt{t^2 -
z^2}$, but more general results are expected in future
investigations.

\bigskip
\section{Acknowledgements}

We wish to thank Gregg Gallatin for interesting conversations.
 The research of AC is
supported in part by DOE grant DE-FG02-92ER-40704. The research of
FC and AS is supported by the DOE. In addition, AC and HM are
supported in part by the Grant-in-Aid for International Scientific
Research No. 09045036, Inter-University Cooperative Research,
Ministry of Education, Science, Sports and Culture of Japan. This
work has been performed as an activity supported by the TMU-Yale
Agreement on Exchange of Scholars and Collaborations. FC and HM are
grateful for the hospitality of the Center for Theoretical Physics
at Yale. HM, AC, and WM are grateful for the hospitality of the
theory group at Los Alamos. In addition, AC wishes to express his
gratitude for the opportunity to present these results at the
TMU-Yale Symposium.

%%%%%%%%%%%%%%%%%%%%%%%%%% Bibliography %%%%%%%%%%%%%%%%%%%%%%%%%%%%%%

\end{document}